\documentclass[%
 reprint,
superscriptaddress,
 amsmath,amssymb,
 aps,
10.5pt]{revtex4-2}
\usepackage{mathrsfs}
\usepackage{siunitx}
\usepackage[dvips]{epsfig}
\usepackage{amsfonts}
\usepackage{xcolor}

\def\rr#1{(\ref{#1})}

\newcommand{\be}{\begin{equation}}
\newcommand{\ee}{\end{equation}}
\usepackage{graphicx}
\usepackage{dcolumn}
\usepackage{bm}

\graphicspath{ {./Chiral_Temperature_Figures} }

\begin{document}

\preprint{APS/123-QED}

\title{{Hydrodynamics of} thermally-driven chiral propulsion and separation}
\author{E. Kirkinis}
\email{kirkinis@northwestern.edu}
\affiliation{Department of Materials Science \& Engineering, Robert R. McCormick School of Engineering and Applied Science, Northwestern University, Evanston IL 60208 USA}
\affiliation{Center for Computation and Theory of Soft Materials, Northwestern University, Evanston IL 60208 USA
}
\author{A. V. Andreev}
\affiliation{Department of Physics, University of Washington, Seattle WA 98195 USA}
\author{M. Olvera de la Cruz}
\affiliation{Department of Materials Science \& Engineering, Robert R. McCormick School of Engineering and Applied Science, Northwestern University, Evanston IL 60208 USA}
\affiliation{Center for Computation and Theory of Soft Materials, Northwestern University, Evanston IL 60208 USA
}

\date{\today}

\begin{abstract}
Considerable effort has been directed towards the characterization of
chiral mesoscale structures, as shown in chiral protein assemblies and carbon nanotubes. 
Here, we establish a thermally-driven hydrodynamic description for the actuation and separation of mesoscale chiral structures in a fluid medium. Cross flow of a Newtonian liquid with a thermal gradient gives rise to
chiral structure propulsion and separation according to their handedness. In turn, the chiral suspension
alters the liquid flow which thus acquires a transverse (chiral) velocity component. Since observation
of the predicted effects requires a low degree of sophistication, our work provides an efficient and inexpensive approach to test and calibrate chiral particle propulsion and separation strategies.
\end{abstract}

\maketitle

{
Chirality, denoting the lack of superposition ability of structures on mirror images, is a characteristic of various  assemblies including carbon nanotubes, viruses and actin filaments, and is essential for their function. 
Since left- and right-handed amino acids lead to different protein structures, their homochirality is required for biological function such as gene encoding \cite{Inaki2016}. 
Chiral proteins can sometimes lead to chiral mesoscale structures; some organisms with chiral body structures have chiral cells \cite{Fan2019}. Therefore, the chirality of proteins may be responsible for the chiral mesoscale structures found in cell media. Chiral mesoscale assemblies are formed in peptide amphiphiles with chiral aminoacids \cite{Gao2019} 
and in many carbon based systems, which have required great efforts to understand and characterize 
\cite{Arnold2006}. However, the mechanism by which chirality manifests at the mesoscale is not well understood.
Here, we propose ways of actuating and separating mesoscale chiral structures
such as helices \cite{Shimizu2005,*McCourt2022}, 
helicoidal scrolls \cite{Nagarsekar2016} and twisted ribbons \cite{Oda1999}.

In chemical and biological systems various mesoscale structures move and function in an aqueous environment in
the presence of thermal gradients induced by chemical reactions \cite{Zhang2014}. Temperature gradients
may alter the liquid material parameters and in particular, viscosity, as this was 
demonstrated in laser-induced thermophoresis experiments 
\cite{schermer2011} and the associated theory of a single hot particle in a viscous liquid \cite{Oppenheimer2016}.  
This motivates us to study the \emph{hydrodynamic} motion of chiral suspensions
in Stokes flow in the presence of temperature gradients. 
The corresponding chiral current $\mathbf{j}^{\textrm{ch}}$ is perpendicular to the 
plane formed by the base
flow direction and the temperature gradient (cf. Fig. \ref{channel}).
The motion of the chiral suspension also perturbs the base flow and endows it with
a transverse (chiral) velocity component. 
It is noteworthy that the chiral suspension also exerts a screw torque on the confining
walls, in the direction of the base flow. 
The hydrodynamic description developed in this article implies averaging over the tumbling motion of the chiral particles and applies at time scales longer than the tumbling time \cite{Andreev2010}. It can be understood as a ``continuum'' formulation for the motion of a chiral suspension and thus \emph{differs} from the majority of propulsion 
descriptions which are based on a resistance matrix at the level of a single
suspended particle \cite{Happel1965}. The equivalence of the two approaches was discussed in the 
recent review article by Witten and Diamant \cite{Witten2020}.  
\begin{figure*}[t]
\vspace{-5pt}
\includegraphics[height=1.6in,width=6.6in]{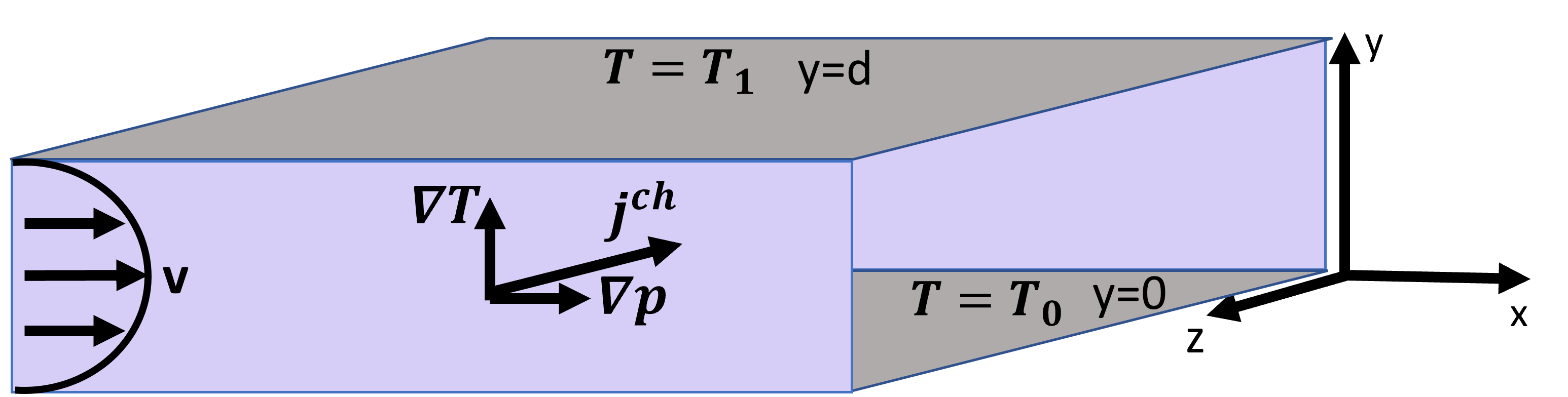}
\vspace{-5pt}
\caption{\label{channel} 
A chiral current ${j}^{\textrm{ch}}\hat{\mathbf{z}}$ of $n$ chiral particles suspended in a classical liquid is induced by a vertical temperature gradient $\nabla T= \partial_y T\hat{\mathbf{y}}$ and shear flow $\mathbf{v} = u(y) \hat{\mathbf{x}}$.
In a nonracemic suspension of chiral particles the liquid is also perturbed by chirality and acquires a transverse chiral velocity component $\delta v(y)\hat{\mathbf{z}}$ perpendicular to the direction of the base flow $u(y) \hat{\mathbf{x}}$.
}
\vspace{-5pt}
\end{figure*}

Motion of chiral particles suspended in a classical liquid is associated with a chiral current
and in particular $\mathbf{j}^\pm = n^\pm \mathbf{v}_p^\pm$
where $n^\pm$ is the number density
of right and left-handed particles, respectively and $\mathbf{v}_p^{\pm}$ is their respective velocity. For simplicity we consider an incompressible liquid where the right and left-handed
particles are mirror-images of each other. We can thus define a chiral current 
$\mathbf{j}^{\textrm{ch}}$ (cf. Fig. \ref{channel}) of the form
\be \label{chiralcurrentdef}
\mathbf{j}^{\textrm{ch}} = \mathbf{j}^+ - \mathbf{j}^-.
\ee 
In the presence of temperature gradients a phenomenological expression for the  chiral current $\mathbf{j}^{\textrm{ch}}$ based on symmetry considerations and which has a low power
of derivatives of vorticity is 
\be \label{j0}
\mathbf{j}^{\textrm{ch}}= \frac{n}{T}\left[\beta_1 (\nabla T \cdot \nabla) \textrm{curl}\mathbf{v}
+ \beta_2 \nabla T \times \nabla^2 \mathbf{v}\right],
\ee
where  $n= n^+  +  n^-$ and $T, \mathbf{v}$ are the liquid's, undisturbed by chirality, 
temperature and velocity respectively. $\beta_{1,2}$ are described below. 
As mentioned above, the magnitude of the chiral current is determined by the complicated tumbling motion of the particles caused by thermal fluctuations and the inhomogeneous flow. The phenomenological expression \rr{j0} is written to lowest order in the driving flow. At strong drives thermal fluctuations are subdominant, and the magnitude of the chiral current is determined by averaging over corresponding Jeffery orbits in the nonuniform flow. This problem, but in a different context, was discussed in \cite{Kirkinis2012a}.  

In Ref. \cite{Andreev2010} the chiral current in an isothermal system was shown to be described by the expression
\be\label{chiralcurrent1}
\mathbf{j}^{\textrm{ch}} = n  \beta \nabla^2 \textrm{curl}\mathbf{v}.
\ee
Insight into the physical origin of Eq.\rr{j0} may be obtained by applying 
Eq.\rr{chiralcurrent1} to a shear flow in the presence of temperature gradients. 
In particular, allowing temperature dependence of the liquid viscosity $\eta$ and implementing the resulting vorticity equation, Eq.\rr{chiralcurrent1} leads to  
\be \label{vorteqn1}
\mathbf{j}^{\textrm{ch}} \sim n\beta \frac{\eta'}{\eta} \left[\nabla T \times \nabla^2 \mathbf{v} +
  (\nabla T \cdot \nabla) \textrm{curl}\mathbf{v} \right],
\ee
where a prime denotes differentiation with respect to temperature $T$ and we
retained only leading order terms in temperature gradients. Thus, the coefficients $\beta_1$ and $\beta_2$
in \rr{j0} can be expressed in terms of the logarithmic derivative of viscosity with respect to temperature. 
We note that in the absence of temperature gradients chiral separation is possible only in nonstationary
or nonlinear flows \cite{Andreev2010} as can be seen by the vorticity equation and \rr{chiralcurrent1}. 
In the presence of temperature gradients chiral separation is possible even in the creeping flow regime. This is important for biological systems, which operate at low Reynolds numbers.

Recent literature examines the way microorganisms and active particles move in liquids with spatially-varying viscosity
\cite{schermer2011, Oppenheimer2016, Datt2019,*Shoele2018}. In what 
follows we are primarily concerned with
suspensions where the viscosity $\eta$ of the base liquid
varies with temperature. A temperature gradient will then give rise to a chiral current of the 
forms \rr{chiralcurrent1} and \rr{vorteqn1}. 

The chiral density $n^{\textrm{ch}} = n^+  -  n^-$ satisfies the conservation law \cite{Andreev2010}
\be \label{conschdensity}
\partial_t n^{\textrm{ch}} + \textrm{div}( \mathbf{v} n^{\textrm{ch}}) + 
\textrm{div}\left[\mathbf{j}(n^{\textrm{ch}}) +\mathbf{j}^{\textrm{ch}}(n)\right] =0,
\ee
where $
\mathbf{j}(n^{\textrm{ch}}) = -D \nabla n^{\textrm{ch}} - n^{\textrm{ch}} \lambda_T \nabla T - n^{\textrm{ch}} \lambda_p \nabla p,
$
is the diffusive current relative to the liquid
\citep{Landau1987,Andreev2010} and $\mathbf{j}^{\textrm{ch}}(n)$ is Eq. \rr{chiralcurrent1}.
The density $n$ of chiral particles satisfies a similar conservation law, which is 
affected by chirality in a non-racemic mixture
\be \label{consn}
\partial_t n + \textrm{div}( \mathbf{v} n) + 
\textrm{div}\left[\mathbf{j}(n) + \mathbf{j}^{\textrm{ch}}(n^{\textrm{ch}}) \right] =0,
\ee
so that Eq. \rr{conschdensity} and \rr{consn} satisfy the Onsager principle of the symmetry of
the kinetic coefficients \cite{Landau1987}.

A chiral suspension imparts stresses on the suspending liquid. To leading order in velocity gradients these stresses, allowed by symmetry, read
\begin{eqnarray} \label{chiralstress1}
\sigma_{ij}^{{ \textrm{ch}}} &=& \eta(T)  
n^{\textrm{ch}}\left\{ \alpha \left[ \partial_i (\textrm{curl}\mathbf{v})_j + \partial_j (\textrm{curl}\mathbf{v})_i \right] \right. \nonumber \\ && +\frac{\alpha_1}{T} 
\left.\left[ \epsilon_{kli} V_{kj} + \epsilon_{klj}  V_{ki} \right]\partial_l T\right\}, 
\label{chiralstress1}
\end{eqnarray}
where $V_{ij}$ is the rate-of-strain tensor. The first term of Eq. \rr{chiralstress1} introduced 
in \cite{Andreev2010} was discussed in the recent review \cite{Witten2020}. The second term
exists only when temperature gradients are present in the liquid. 

The coefficients $\beta$, $\alpha$ and $\alpha_1$ in \rr{chiralcurrent1} and \rr{chiralstress1} are determined in the low
Reynolds number regime by studying the particle motion in the surrounding liquid \citep{Happel1965}. They
may be estimated as 
\be \label{alphabeta}
\alpha \sim\alpha_1\sim \chi R^4 \quad \textrm{and} \quad \beta \sim \chi R^3, 
\ee
where $R$ is the chiral particle radius and $\chi$ is the degree of chirality in the shape of the particles. 
Eq. \rr{alphabeta} provides the order of magnitude estimates of these coefficients. Their precise determination for a specific particle shape however requires solving hydrodynamic equations for a tumbling particle in the presence of temperature and velocity gradients, and is beyond the scope of our work. 

In the \emph{absence} of chirality
the liquid satisfies the Navier-Stokes equations and is considered incompressible
\be\label{indexeqn}
\rho {D u_i}/{D t} = \partial_k \sigma_{ik} \quad \textrm{and} \quad \partial_i u_i =0, 
\ee
where the Cauchy stress tensor $\sigma_{ik}$ is given by 
$
\sigma_{ik} = -p\delta_{ik} +\eta \left( \frac{\partial u_i}{\partial x_k} + \frac{\partial u_k}{\partial x_i}
\right)i,k = 1,2,3,
$
$\rho$ the density of the liquid and $p$ is the pressure. 
Conservation of energy in an incompressible liquid is expressed in the form \citep{Landau1987} 
\be \label{LL50.1a}
\rho c_p (\partial_t T+ \mathbf{v}\cdot \textrm{grad} T) = k_{th} \nabla^2 T
\ee
where
$c_p$ is the specific heat at constant pressure and $k_{th}$ the thermal conductivity of the liquid.

Consider pressure-driven flow, in the absence of chiral particles, in a channel with unevenly heated walls (cf. Fig. \ref{channel}). With
$\mathbf{v} = u(y) \hat{\mathbf{x}}$, $\nabla T= \partial_y T\hat{\mathbf{y}}$, 
the Navier-Stokes equations \rr{indexeqn} and energy balance \rr{LL50.1a} in the creeping flow approximation reduce to 
\be\label{ns1}
\frac{d}{dy}\left[\eta(T) \frac{du}{dy}\right] = \frac{dp}{dx}, \qquad \frac{d^2 T}{dy^2} = 0, 
\ee
respectively, with boundary conditions 
\be \label{nsbc1}
u(0) = u(d) = 0, \quad T(0) =T_0, \quad T(d) = T_0 + \Delta T. 
\ee
The temperature profile thus obtained is 
$
T (y) = T_0 + \frac{y}{d} \Delta T$. 
The solution of the first of Eq. \rr{ns1} with
boundary conditions \rr{nsbc1} becomes
\begin{widetext}
\begin{equation} \label{uP}
u(y) = \frac{T_e^2d^2 \partial_x p}{2\eta(T) (\Delta T)^2}
\frac{\sum\limits_{\left\{i,j,k\right\}} e^{X_{i}}\left\{\textrm{Ei}_1(X_i) \left[ \frac{e^{X_{j}}}{X_k^2} - 
\frac{e^{X_{k}}}{X_j^2} \right] +  \frac{1}{X_jX_k}
\left(\frac{1}{X_k} - \frac{1}{X_j} \right)\right\}}
{\left[ \textrm{Ei}_1(X_0) - \textrm{Ei}_1(X_1) \right] e^{X_{0} + X_{1}} +  \frac{e^{X_{0}}}{X_1} -  \frac{e^{X_{1}}}{X_0}  },
\end{equation}
\end{widetext}
where the symbol $\left\{i,j,k\right\}$ denotes cyclic permutation of $i, j$ and $k$, and 
$
\textrm{Ei}_1(X) =  \int_1^\infty \frac{e^{-k X}}{k} dk
$ 
is the exponential integral. 
Here we employed the well-documented Arrhenius-type law \citep{Fogelson2001}
\be \label{Arr}
\eta(T) = \eta_0 e^{\frac{E}{R_g(T+T_A)}}, 
\ee
valid in the $243$ to $373$ K temperature range, since it encompasses linear and other exponential 
laws \citep{Potter1972,*Davis1983,*Wall1996}, as special cases. 
$E$ is the activation energy, $R_g$ is the gas constant and $T_A$ is a temperature correction, 
unique to each viscous liquid, cf. \citep{Fogelson2001} and Table \ref{table: table1}.  
In Eq. \rr{uP}
$
X_i = \frac{Te}{T_A+ T_i}, \quad
i= 0,1,2, \quad T_e = \frac{E}{R_g},
$ 
where $T_0$ and $T_1=T_0 +\Delta T$ are the lower and upper channel wall fixed temperatures, respectively
(cf. Fig. \ref{channel}) and
$T_2 \equiv T= T_0 + \frac{y}{d} \Delta T$.

{Now consider the presence of chiral particles and define the chiral separation velocity
\be
\mathbf{v}^{\textrm{ch}} \equiv \mathbf{j}^{\textrm{ch}}/n
\ee
relative to the liquid by employing \rr{chiralcurrent1}. With respect to the geometry displayed in Fig. \ref{channel} it has the form
$\mathbf{v}^{\textrm{ch}} = v^{\textrm{ch}}(y) \hat{\mathbf{z}}$ and its
magnitude is}
\begin{widetext}
\be \label{vpexact}
v^{\textrm{ch}}(y) = \chi \frac{R^3}{d} \frac{X_2^4 \Delta T \partial_xp}{2\eta(T)  T_e}
\frac{\left[ \textrm{Ei}_1(X_1) - \textrm{Ei}_1(X_0) \right] e^{X_{0} + X_{1}} +\sum\limits_{i\neq j =0,1} \frac{(-1)^i e^{X_{j}}}{X_i^2} \left[\frac{1}{2}+\frac{1}{X_i}\left(\frac{1}{X_2}-\frac{1}{2}\right) \right]  }
{\left[ \textrm{Ei}_1(X_0) - \textrm{Ei}_1(X_1) \right] e^{X_{0} + X_{1}} +  \frac{e^{X_{0}}}{X_1} -  \frac{e^{X_{1}}}{X_0}}.
\ee
\end{widetext} 
In Fig. \ref{vp_y} we plot the closed form expression \rr{vpexact} for the 
chiral separation velocity $v^{\textrm{ch}}$ in cm/sec vs. channel elevation $y$ in cm for two temperature
variations $\Delta T$ between the lower (at $y=0$) and upper (at $y=0.1$ cm) channel walls. 
The chiral separation velocity is non-zero close to the solid walls located at $y=0,d$, even though
no-slip boundary conditions are satisfied by the base liquid. 
This is the case because, according to Eq. \rr{chiralcurrent1}, chiral particle velocities
become prominent in the vicinity of large vorticity gradients, and these are present close to the
walls.
\begin{table}[t]
\caption{\label{table: table1}%
Definitions and material parameters \cite{Fogelson2001}}
\begin{ruledtabular}
\begin{tabular}{lcl}
\textrm{Quantity}&
\textrm{Value}&
\textrm{Definition}\\
\colrule
$\eta$ ($\textrm{g}\:\textrm{cm}^{-1}\textrm{sec}^{-1} $)    & 2.03         & viscosity of BM-4 oil at \SI{25}{\celsius} \cite{Fogelson2001}\\
$R$ (cm)  & $5\times 10^{-3}$& chiral particle radius\\
$d$ (cm) & $0.1$ & channel width\\
$U_0$ (cm/sec) & $0.1$ & Poiseuille velocity\\
$T_0$ (K) & $298.15$& lower channel wall temperature\\
$\gamma$ ($K^{-1}$) &  $0.07$ & BM-4 oil \cite{Fogelson2001}\\
$E$ (kJ/mol) & $7.5$& activation energy of BM-4 oil \cite{Fogelson2001}\\
$R_g$ (J/ (mol K) & $8.31441$& gas constant\\
$T_{\textrm{A}}$ (K) & $-186$& BM-4 oil temp. correction \cite{Fogelson2001}\\
$n^{\textrm{ch}}  (\textrm{cm}^{-3})$ & $R^{-3}$ & chiral density $n^+-n^-$\\
$n\; (\textrm{cm}^{-3}) $ & $R^{-3}$ & particle number density $n^++n^-$\\
$u$ (cm/sec) & &basic shear flow velocity\\
$v^{\textrm{ch}}$ (cm/sec)& & chiral separation velocity\\
$\delta v$ (cm/sec)& &chiral correction to flow velocity
\end{tabular}
\end{ruledtabular}
\end{table}

To obtain a better understanding of the effect, we 
average \rr{vpexact} over the channel width $d$ and expand with respect to $\Delta T$ to obtain 
$\langle v^{\textrm{ch}} \rangle \sim 2\chi R^3 \gamma \frac{\Delta T}{d}\frac{\partial_x p}{\eta}$, to leading order
in $\Delta T$. It is more illuminating 
however to replace the pressure gradient with a characteristic velocity $U_0$ of Poiseuille flow
by averaging the (undisturbed by chirality) base Poiseuille profile $u \sim \frac{\partial_x p}{2\eta(T_0)}(y^2 - yd)$ over the channel width $d$. This gives
$U_0 = - \frac{\partial_x p}{12\eta(T_0)}d^2$, and substituting into the expression for 
$\langle v^{\textrm{ch}} \rangle$ we obtain the 
chiral separation velocity
\be \label{singlevp0}
\langle v^{\textrm{ch}} \rangle = 24 \chi \left( \frac{R}{d} \right)^3 U_0 \gamma \Delta T,
\ee
where we defined the average of a function $f(y)$ with respect to the channel width to be
$\langle f \rangle = \frac{1}{d} \int_0^d f(y) dy$, $R$ is chiral particle size and $d$ channel width. 
$\gamma = {E}/\left[{R_g(T_0 + T_A)^2}\right]$ is the logarithmic derivative of the viscosity 
which arises from the linearization of \rr{Arr}
$
\eta (T) \sim \eta(T_0) \left[ 1 - \gamma (T-T_0)\right]. 
$

Considering BM-4 oil \citep{Fogelson2001}, $\Delta T = \SI{10}{\kelvin}$ and the values displayed in Table \ref{table: table1}, Eq. \rr{singlevp0} leads to the estimate
\be\label{singlevpest}
v^{\textrm{ch}} \sim 2\chi  \;\mu\textrm{m/sec}. 
\ee
The Reynolds number is $\textrm{Re} \sim 3.4\times 10^{-3}$. Analogous results can be derived for silicon oils employed in the experiments of Ehrhard \cite{Ehrhard1993} and other liquids reported in the literature \citep{Fogelson2001}. Water can also be used, although it leads to Reynolds numbers higher than those reported here.

\emph{Pertubation of liquid velocity by the chiral suspension.}-
A chiral suspension imparts stresses on the suspending liquid. To leading order in gradients of vorticity these stresses, allowed by symmetry 
are given by \rr{chiralstress1}.
The liquid velocity $\mathbf{v}$ acquires a chirality-induced component
$\delta v$ perpendicular to the plane of the base flow
\be
\mathbf{v} = u(y)\hat{\mathbf{x}}+\delta v(y)\hat{\mathbf{z}}. 
\ee
With the correction \rr{chiralstress1} the Cauchy stress tensor reads
\be
\sigma_{ij} = -p \delta_{ij} + \eta (\partial_i u_j + \partial_ju_i) +\sigma_{ij}^{{ \textrm{ch}}}. 
\ee
Conservation of linear momentum $\partial_j \sigma_{ij} = 0$ along the flow direction
$
\hat{\mathbf{x}}:    -\partial_xp  +\partial_y (\eta \partial_yu)=0 ,
$
is now accompanied by its chirality-induced counterpart that is perpendicular to the base flow direction
\be \label{cheqn1}
\hat{\mathbf{z}}: \partial_y( \eta \partial_y \delta v) - n^{\textrm{ch}} \chi R^4\partial_y(\eta \partial_y^2 u) = 0,
\ee
and satisfies no-slip boundary conditions
$
\delta v(0) =   \delta v(d) = 0. 
$
\begin{figure}
\vspace{-5pt}
\begin{center}
\includegraphics[height=2.8in,width=3.4in]{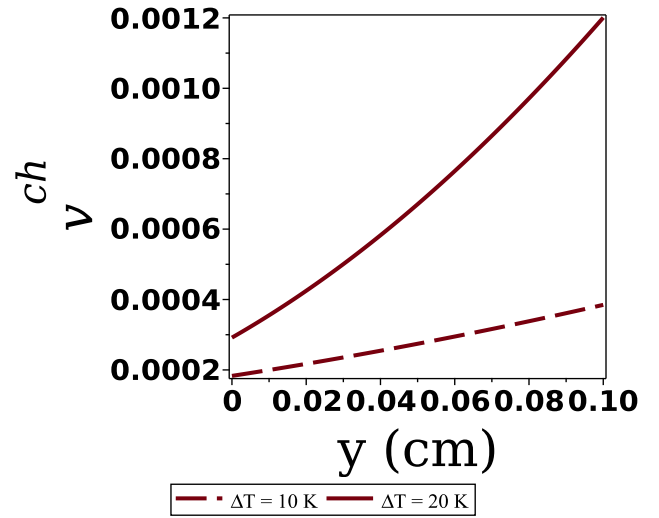}
\end{center}
\caption{\label{vp_y} 
Chiral separation velocity $v^{\textrm{ch}}$ in cm/sec from Eq. \rr{vpexact}, perpendicular to the $x$-$y$ plane formed by a base Poiseuille flow and the vertical temperature gradient
directions (cf. Fig.\ref{channel}). $y$ is vertical channel coordinate in cm and we employed the Arrhenius-type temperature dependent viscosity law \rr{Arr} for a BM-4 oil \citep{Fogelson2001}. $\chi$ has been set equal to $1$.}
\end{figure}
The solution $\delta v$ of Eq. \rr{cheqn1} is displayed in Fig. \ref{vch_y} in cm/sec vs. channel elevation $y$ in cm for two temperature
variations $\Delta T$ between the lower (at $y=0$) and upper (at $y=0.1$ cm)
channel walls, employing the Arrhenius-type temperature dependent viscosity law \rr{Arr}. Its profile is skewed 
due to the reduction of viscosity close to the upper heated channel wall which is also the location
of high chiral separation velocity $v^{\textrm{ch}}$. 
To leading order in $\Delta T$, and averaging over the channel width $d$, we obtain 
$\langle \delta v \rangle \sim \chi R \frac{d\partial_x p}{12\eta} \gamma \Delta T$. Replacing the pressure gradient with its Poiseuille flow counterpart, leads to 
\be
\langle \delta v \rangle \sim \chi \frac{R}{d} U_0 \gamma \Delta T.
\ee
Employing the material parameters for BM-4 oil displayed in Table \ref{table: table1} and setting 
$\Delta T = \SI{10}{\kelvin}$ we obtain $
\delta v \sim 35\chi  \;\mu\textrm{m/sec}, 
$
which agrees well, in order of magnitude, with the exact solution displayed in Fig. \ref{vch_y}. 
The momentum equation displayed in \rr{cheqn1} was formulated by 
considering only the first term of the constitutive law \rr{chiralstress1} since the second term, and for the material parameters
employed in this article, gives velocities that are one order
of magnitude smaller than the ones derived here.

\begin{figure}[t]
\begin{center}
\includegraphics[height=2.8in,width=3.4in]{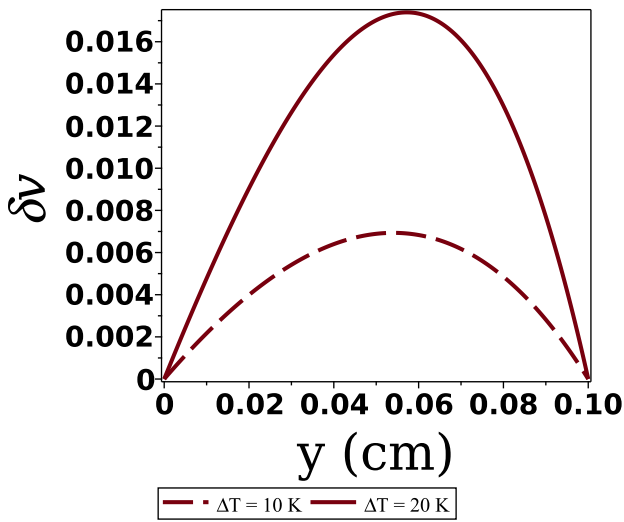}
\end{center}
\caption{\label{vch_y} 
Transverse chiral component of liquid velocity $\delta v$ in cm/sec, perpendicular to the $x$-$y$ plane formed by a Poiseuille flow and the temperature gradient
directions (cf. Fig.\ref{channel}) and obtained by solution of Eq. \rr{cheqn1} with boundary conditions $
\delta v(0) =   \delta v(d=0.1) = 0
$. $y$ is vertical channel coordinate in cm and we employed the Arrhenius-type temperature dependent viscosity law \rr{Arr} for a BM-4 oil \cite{Fogelson2001} giving
rise to the skewness of chiral velocity profiles. $\chi$ has been set equal to $1$.}
\end{figure}

\emph{Screw torque in a non-racemic suspension.}-
A non-racemic mixture will apply shear stresses on the channel walls that are perpendicular to the plane
of the paper. These forces arise from the chiral momentum flux density \rr{chiralstress1},
cf. \citep{Andreev2010}.
Employing the geometry of the channel Poiseuille flow displayed in Fig. \ref{channel}
this stress is 
\be \label{chstressP}
\hat{\mathbf{z}}: \quad \sigma_{zy}^{{ \textrm{ch}}} = \chi R \eta \partial^2_y u. 
\ee
In Fig. \ref{sigmach_y} we display the chiral stress $\sigma_{zy}^{{ \textrm{ch}}}$ as a function
of channel width employing the exact form for the liquid velocity profile \rr{uP}. Since the normal vectors to the two channel walls have opposite sign, the chiral suspension 
exerts on the walls two forces of opposite sign directed into and out of the page. 
Hence, there is a screw torque exerted by the chiral flow on the confining walls and is directed along the
$\hat{\mathbf{x}}$ direction of the flow. 
The average of the chiral stress
$\langle  \sigma_{zy}^{{ \textrm{ch}}} \rangle$ over the channel width to leading order in $\Delta T$
becomes
\be \label{szych}
\langle  \sigma_{zy}^{{ \textrm{ch}}} \rangle = \chi R \partial_x p (1 + \frac{1}{2}\gamma \Delta T
+ O((\Delta T)^2).
\ee
Expression \rr{szych} implies that a chiral stress exists even in the absence of temperature gradients.
This was also noted in \citep{Andreev2010}. 
Replacing the pressure gradient by the base Poiseuille profile, as carried out in the foregoing sections and employing
the material values appearing in Table \ref{table: table1} for $\Delta T = 10 K$, Eq.\rr{szych} gives
$
\langle  \sigma_{zy}^{{ \textrm{ch}}} \rangle  = 1.73\chi \textrm{ dynes/cm}^{-2}. $
This estimate agrees, in order of magnitude, with those of same-size systems known in the literature  \cite{Kataoka1999,*Ando2009}.

\begin{figure}[t]
\vspace{5pt}
\begin{center}
\includegraphics[height=2.6in,width=3.2in]{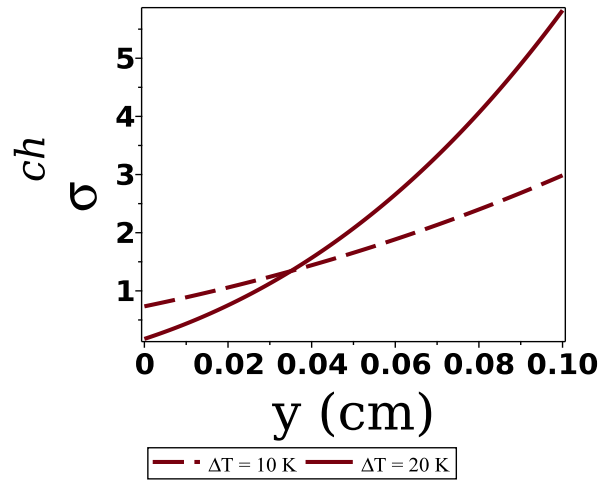}
\end{center}
\caption{\label{sigmach_y} 
Distribution of the chiral stress $\sigma_{zy}^{{ \textrm{ch}}}$ in \rr{chstressP} in dynes/$\textrm{cm}^2$ imparted by the chiral suspension on the liquid in a direction perpendicular to the $x$-$y$ plane formed by the Poiseuille flow and the vertical temperature gradient (cf. Fig.\ref{channel}), vs. the vertical coordinate $y$ of the channel in cm and we employed the Arrhenius-type temperature dependent viscosity law \rr{Arr} for a BM-4 oil \cite{Fogelson2001}. 
Since the normal vectors to the two channel walls have opposite signs, the chiral suspension 
exerts on the walls two forces of opposite sign directed into and out of the page. 
Hence, there is a screw torque exerted by the chiral flow on the confining walls and is directed along the
base flow direction. 
$\chi$ has been set equal to $1$.}
\end{figure}

We note the existence of a related thermal effect for the propulsion of chiral particles 
when the viscosity is temperature-dependent and
thermal gradients are generated at the interior by viscous heating \cite{Kirkinis2019a}. 
The energy equation is now replaced by $
\kappa \partial_y^2 T + \eta(T)\left(\partial_y u\right)^2 =0$, where $\kappa$ is the thermal conductivity
of the liquid.
The external stimuli may be supplied, for instance, by sliding one channel wall at constant speed $V$.
In this case the vorticity equation and the chiral current depend on the Brinkman number $Br$, that is, 
the ratio of viscous heating to conduction $Br = \frac{\gamma \eta V^2}{\kappa}$. The 
chiral separation velocity becomes
$
v^{\textrm{ch}}  \sim \frac{\chi}{8} \left(\frac{R}{d}\right)^3 Br. 
$

Another related thermally-induced chiral particle propulsion effect
can take place in a Rayleigh-B\`enard cell \cite{Chandrasekhar1961}, driven by 
variations of the liquid density with temperature. Here, the coefficient of thermal expansion $\alpha_T$ is the logarithmic derivative of density with respect to temperature, in the same sense that $\gamma$
in Eq.~\rr{singlevp0}
is the logarithmic derivative of viscosity.
Diffusion of vorticity, perpendicular to the plane of the cell, is now induced by 
$\rho \alpha_T \nabla T \times \mathbf{g}$, 
where
$\rho$ is the unperturbed mass density of the liquid
and both the temperature gradient and
gravitational acceleration $\mathbf{g}$ lie on the plane of the cell. The chiral separation velocity is
$ \label{singleRB}
v^{\textrm{ch}}  \sim \chi \frac{R^3 \rho g \alpha_T  \Delta T}{\eta d}.
$
This effect is present when the Rayleigh number exceeds its critical value.

We thank the Department of Energy, Office of
Basic Energy Sciences for support under contract DE-FG02-08ER46539 (M.O.C and E.K.).
The work of A.V.A. was supported, in part, by the US National Science Foundation through the MRSEC Grant No. DMR-1719797.

\bibliographystyle{apsrev4-2}
\bibliography{/Users/eks400/Documents/Bibliography/disorder,/Users/eks400/Documents/Bibliography/materialscience,/Users/eks400/Documents/Bibliography/perturbations,/Users/eks400/Documents/Bibliography/fluids,/Users/eks400/Documents/Bibliography/physiology,/Users/eks400/Documents/Bibliography/Compressible}

\end{document}